\documentclass{midl}

\usepackage{mwe} 

\usepackage{hyperref}
\usepackage{amsmath,amsfonts,amssymb}
\usepackage{graphicx}
\usepackage{times}
\usepackage{amsmath}
\usepackage{nccmath}
\usepackage{amssymb}
\usepackage{mwe}
\usepackage{amssymb}
\usepackage{xcolor,colortbl}
\usepackage{relsize}
\usepackage{pifont}
\usepackage{booktabs} 
\usepackage{adjustbox}
\usepackage{multirow}
\usepackage{multicol}
\usepackage{nicematrix}
\usepackage{float}
\usepackage{graphicx}
\usepackage{bm}
\usepackage{makecell}
\usepackage{tabu}

\jmlrproceedings{MIDL}{Medical Imaging with Deep Learning}
\jmlrpages{}
\jmlryear{2023}

\jmlrworkshop{Short Paper -- MIDL 2023 submission}
\jmlrvolume{-- Under Review}
\editors{Under Review for MIDL 2023}

\title[Short Title]{Segment Anything Model (SAM) for Digital Pathology: \\ Assess Zero-shot Segmentation on Whole Slide Imaging}

\midlauthor{\Name{Ruining Deng\midljointauthortext{Joint first author: contributed equally}\nametag{$^{1}$}} \Email{r.deng@vanderbilt.edu}\\
\Name{Can Cui\midlotherjointauthor\nametag{$^{1}$}} \Email{can.cui.1@vanderbilt.edu}\\
\Name{Quan Liu\midlotherjointauthor\nametag{$^{1}$}} \Email{quan.liu@vanderbilt.edu}\\
\Name{Tianyuan Yao\nametag{$^{1}$}} \Email{tianyuan.yao@vanderbilt.edu}\\
\Name{Lucas W. Remedios\nametag{$^{1}$}}
\Email{lucas.w.remedios@vanderbilt.edu}\\
\Name{Shunxing Bao\nametag{$^{1}$}}
\Email{shunxing.bao@vanderbilt.edu}\\
\Name{Bennett A. Landman\nametag{$^{1}$}} \Email{bennett.landman@vanderbilt.edu}\\
\Name{Lee E. Wheless\nametag{$^{2,3}$}} 
\Email{lee.e.wheless@vumc.org}\\
\Name{Lori A. Coburn\nametag{$^{2,3}$}} 
\Email{lori.coburn@vumc.org}\\
\Name{Keith T. Wilson\nametag{$^{2,3}$}} \Email{keith.wilson@vumc.org}\\
\Name{Yaohong Wang\nametag{$^{2}$}} \Email{yaohong.wang@vumc.org}\\
\Name{Shilin Zhao\nametag{$^{2}$}} \Email{shilin.zhao.1@vumc.org}\\
\Name{Agnes B. Fogo\nametag{$^{2}$}} \Email{agnes.fogo@vumc.org}\\
\Name{Haichun Yang\nametag{$^{2}$}} \Email{haichun.yang@vumc.org}\\
\Name{Yucheng Tang\nametag{$^{4}$}} \Email{yuchengt@nvidia.com}\\
\Name{Yuankai Huo\midljointauthortext{Corresponding author}\nametag{$^{1}$}} \Email{Yuankai.huo@vanderbilt.edu}\\
\addr $^{1}$ Vanderbilt University, Nashville, TN, USA  \\
\addr $^{2}$ Vanderbilt University Medical Center, Nashville, TN, USA \\
\addr $^{3}$ Veterans Affairs Tennessee Valley Healthcare System, Nashville, TN, USA \\
\addr $^{4}$ NVIDIA Cooperation, Redmond, WA, USA}

\begin{document}

\maketitle

\begin{abstract}
The segment anything model (SAM) was released as a foundation model for image segmentation. The promptable segmentation model was trained by over 1 billion masks on 11M licensed and privacy-respecting images. The model supports zero-shot image segmentation with various segmentation prompts (e.g., points, boxes, masks). It makes the SAM attractive for medical image analysis, especially for digital pathology where the training data are rare. In this study, we evaluate the zero-shot segmentation performance of SAM model on representative segmentation tasks on whole slide imaging (WSI), including (1) tumor segmentation, (2) non-tumor tissue segmentation, (3) cell nuclei segmentation. \textbf{Core Results:} \textit{The results suggest that the zero-shot SAM model achieves remarkable segmentation performance for large connected objects. However, it does not consistently achieve satisfying performance for dense instance object segmentation, even with 20 prompts (clicks/boxes) on each image.} We also summarized the identified limitations for digital pathology: (1) image resolution, (2) multiple scales, (3) prompt selection, and (4) model fine-tuning. In the future, the few-shot fine-tuning with images from downstream pathological segmentation tasks might help the model to achieve better performance in dense object segmentation. 
\end{abstract}


\section{Introduction}

Large language models (e.g., ChatGPT~\cite{brown2020language} and GPT-4~\cite{openai2023gpt4}), are leading a paradigm shift in natural language processing with strong zero-shot and few-shot generalization capabilities. This development has encouraged researchers to develop large-scale vision foundation models. While the first successful "foundation models"~\cite{bommasani2021opportunities} in computer vision have focused on pre-training approaches (e.g., CLIP~\cite{radford2021learning} and ALIGN~\cite{jia2021scaling}) and generative AI applications (e.g., DALL·E~\cite{ramesh2021zero}), they have not been specifically designed for image segmentation tasks~\cite{kirillov2023segment}. Segmenting objects (e.g., tumor, tissue, cell nuclei) for whole slide imaging (WSI) data is an essential task for digital pathology, deep learning models typically necessitate well-delineated training data. Obtaining these gold-standard data from clinical experts can be challenging due to privacy regulations, intensive manual efforts, insufficient reproducibility, and complicated annotation processes~\cite{huo2021ai}. Hence, zero-shot image segmentation~\cite{wang2019survey} is desired, where the model can accurately segment pathological images without prior exposure to the domain data during training.

Recently, the "Segment Anything Model" (SAM)~\cite{kirillov2023segment} was proposed as a foundation model for image segmentation. The model has been trained on over 1 billion masks on 11 million licensed and privacy-respecting images. Furthermore, the model supports zero-shot image segmentation with various segmentation prompts (e.g., points, boxes, and masks). This feature makes it particularly attractive for pathological image analysis where the labeled training data are rare and expensive.

In this study, we assess the zero-shot segmentation performance of the SAM model on representative segmentation tasks, including (1) tumor segmentation~\cite{liu2021simtriplet}, (2) tissue segmentation~\cite{deng2023omni},  and (3) cell nuclei segmentation~\cite{li2021beds}. Our study reveals that the SAM model has some limitations and performance gaps compared to state-of-the-art (SOTA) domain-specific models.

\section{Experiments and Performance}
We obtained the source code and the trained model from \url{https://segment-anything.com}. To ensure scalable assessments, all experiments were performed directly using Python, rather than relying on the Demo website. The results are presented in Figure \textcolor{green}{1} and Table \textcolor{green}{1}.

\begin{figure}[t]
\begin{center}
\includegraphics[width=0.95\linewidth]{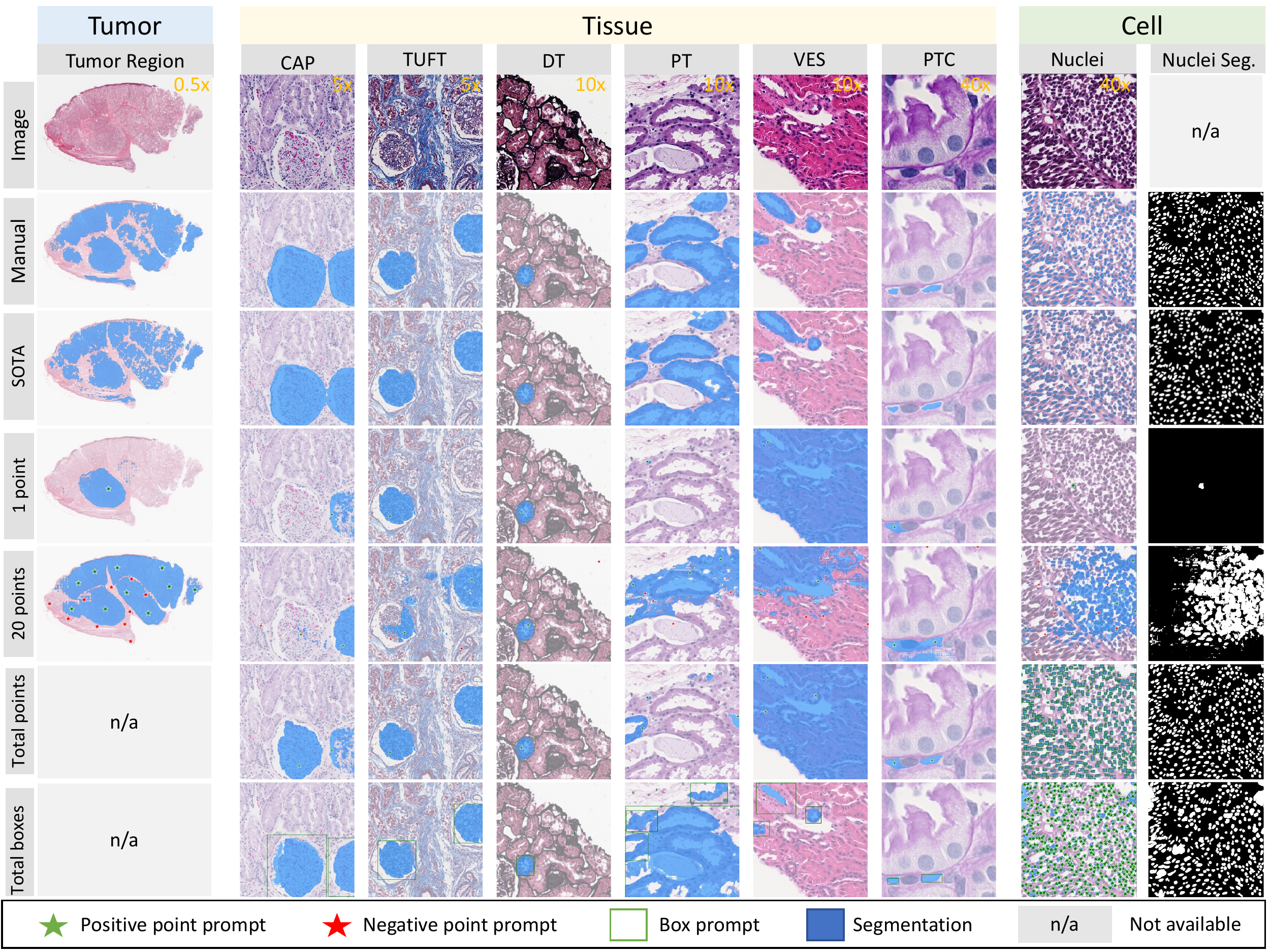}
\end{center}
   \caption{\textbf{Qualitative segmentation results}. The SOTA methods are compared with SAM method with different prompt strategies. }
\label{fig:tumor_seg}
 \end{figure}

\begin{table}[bht]
 \caption{Compare SAM with state-of-the-art (SOTA) methods. (Unit: Dice score)}
  \centering
  \begin{tabular*}{\textwidth}{@{\extracolsep{\fill}}lccccccccc}
    \toprule

\multirow{3}{*}{Method}   &  \multirow{3}{*}{Prompts} & Tumor & \multicolumn{6}{c}{Tissue}  & Cell \\

\cmidrule(lr){3-3} 
\cmidrule(lr){4-9}
\cmidrule(lr){10-10}

  &  & 0.5$\times$ & \multicolumn{2}{c}{5$\times$} & \multicolumn{3}{c}{10$\times$} & \multicolumn{1}{c}{40$\times$}  & 40$\times$ \\

\cmidrule(lr){3-3} 
\cmidrule(lr){4-5} 
\cmidrule(lr){6-8} 
\cmidrule(lr){9-9} 
\cmidrule(lr){10-10}

 & & Tumor & CAP & TUFT & DT  & PT  & VES & PTC   & Nuclei \\
    \midrule
    SOTA   &  no prompt & 71.98 & \textcolor{red}{96.50} & \textcolor{red}{96.59} & \textcolor{red}{81.01} & \textcolor{red}{89.80} & \textcolor{red}{85.05} & \textcolor{red}{77.23}  &   \textcolor{red}{81.77} \\

    SAM &  1 point & 58.71 & 78.08 & 80.11 & 58.93 & 49.72 & 65.26 & 67.03  & 1.95 \\ 
    
      SAM  &  20 points & \textcolor{red}{74.98}  & 80.12 & 79.92 & 60.35 & 66.57 & 68.51 & 64.63  & 41.65\\

    \cline{2-10}
    
      SAM   &  \cellcolor[gray]{0.9}total points &\cellcolor[gray]{0.9} n/a  &\cellcolor[gray]{0.9} 88.10 &\cellcolor[gray]{0.9} 89.65 &\cellcolor[gray]{0.9} 70.21 &\cellcolor[gray]{0.9} 73.19 &\cellcolor[gray]{0.9} 67.04 &\cellcolor[gray]{0.9} 67.61 &\cellcolor[gray]{0.9} 69.50 \\ 
    
       SAM  &  \cellcolor[gray]{0.9}total boxes &\cellcolor[gray]{0.9} n/a   &\cellcolor[gray]{0.9} 95.23 &\cellcolor[gray]{0.9} 96.49 &\cellcolor[gray]{0.9} 89.97 &\cellcolor[gray]{0.9} 86.77 &\cellcolor[gray]{0.9} 87.44 &\cellcolor[gray]{0.9} 87.18 &\cellcolor[gray]{0.9} 88.30  \\  
    \bottomrule
  \end{tabular*}
    {total points/boxes: we place points/boxes on every single instance object (based on the known ground truth) as a theoretical upper bound of SAM. Note that it is impractical in real applications.}
\end{table}

\textbf{Tumor Segmentation}.The whole-slide images (WSIs) of skin cancer patients were obtained from the Cancer Genome Atlas (TCGA) datasets (TCGA Research Network: \url{https://www.cancer.gov/tcga}). We employed SimTriplet~\cite{liu2021simtriplet} approach as the SOTA method, with the same testing cohort to make a fair comparison. In order to be compatible with the SAM segmentation model, the WSI inputs were scaled down 80 times from a resolution of 40$\times$, resulting in an average size of 860$\times$1279 pixels. We evaluated two different scenarios: (1) SAM with a single positive point prompt, and (2) SAM with 20 point prompts (10 positive and 10 negative points). The prompts were randomly selected from manual annotations, with positive prompt points being selected from the tumor region and negative prompt points being selected from the non-tumor region.

\textbf{Tissue Segmentation}. 
A total of 1,751 regions of interest (ROIs) images were obtained from 459 WSIs belonging to 125 patients diagnosed with Minimal Change Diseases. These images were manually segmented to identify six structurally normal pathological primitives~\cite{jayapandian2021development}, using digital renal biopsies from the NEPTUNE study~\cite{barisoni2013digital}. To form a test cohort for multi-tissue segmentation, we captured 8,359 patches measuring 256$\times$256 pixels. For comparison, We employed Omni-Seg~\cite{deng2023omni} approach as the SOTA method. The tissue types consist of the  glomerular unit (CAP), glomerular tuft (TUFT), distal tubular (DT), proximal tubular (PT), arteries (VES), and peritubular capillaries (PTC). For the SAM method, we evaluated four different scenarios: (1) SAM with a single positive point prompt, (2) SAM with 20 point prompts (10 positive and 10 negative points), and (3)/(4) SAM with all points/boxes on every single instance object, which served as a theoretical upper bound for SAM. We randomly selected point prompts from the manual annotations and eroded each connected component with a 10x10 filter to generate at most one random point. For the box prompts, we used the bounding box of each connected component.

\textbf{Cell nuclei Segmentation}.
The dataset for nuclei segmentation was obtained from the MoNuSeg challenge~\cite{kumar2019multi}. It contains H\&E stained images at 40$\times$ magnification with 1000$\times$1000 pixels from the TCGA dataset, along with corresponding annotations of nuclear boundaries. The MoNuSeg dataset includes 30 images for training and 14 for testing. We evaluated the performance of SAM models against the BEDs model~\cite{li2021beds}, a competitive nuclei segmentation model trained on the MoNuSeg training data. The prompt method and evaluation are as described in $\mathsection$\textbf{Tissue Segmentation}.

\section{Limitations on Digital Pathology}
The SAM models achieve remarkable performance under zero-shot learning scenarios. However, we identified several limitations during our assessment.

\textbf{Image resolution}. The average training image resolution of SAM is 3300$\times$4950 pixels~\cite{kirillov2023segment}, which is significantly smaller than Giga-pixel WSI data ($>10^9$ pixels). Moreover, analyzing WSI data at the patch level may result in an impractical number of interactions, even if only a small number of points or bounding boxes are marked for each patch.

\textbf{Multiple scales}. Multi-scale is a significant feature in digital pathology. Different tissue types have their optimal image resolution (as shown in Table \textcolor{green}{1}). For instance, at the optimal resolution for CAP segmentation (5$\times$ scale), it is difficult to achieve good segmentation for PTC. However, zooming in (40$\times$ scale) would result in nearly 100 times more patches.

\textbf{Prompt selection}. Firstly, to achieve decent segmentation performance in zero-shot learning scenarios, a considerable number of prompts are still necessary. Secondly, the segmentation performance heavily depends on the quality of prompt selection. Another concern related to segmentation performance is inter-rater and intra-rater reproducibility of prompt-based segmentation.

\textbf{Model fune-tuning}. Currently, tedious manual prompt placements are still necessary for segmentation tasks with significant domain heterogeneities. A reasonable online/offline fine-tuning strategy is necessary to propagate the knowledge obtained from manual prompts to larger-scale automatic segmentation on Giga-pixel WSI data.

\section{Conclusion}
The zero-shot setting of SAM enables domain users to segment heterogeneous objects in digital pathology without undergoing a heavy training process. The results suggest that the zero-shot SAM model achieves remarkable segmentation performance for large connected objects. However, it
does not consistently achieve satisfying performance for dense instance object segmentation, even
with 20 prompts (clicks/boxes) on each image. Nonetheless, several limitations still exist and require further investigation for digital pathology.

\midlacknowledgments{This research was supported by NIH R01DK135597 (Huo), The Leona M. and Harry B. Helmsley Charitable Trust grant G-1903-03793 and G-2103-05128, NSF CAREER 1452485, NSF 2040462, NCRR Grant UL1 RR024975-01 (now at NCATS Grant 2 UL1 TR000445-06), NIH NIDDK DK56942 (ABF), DoD HT94252310003 (Yang), the VA grants I01BX004366 and I01CX002171, VUMC Digestive Disease Research Center supported by NIH grant P30DK058404, NVIDIA hardware grant, resources of ACCRE at Vanderbilt University.}

\bibliography{midl-samplebibliography}

\end{document}